\documentclass[reprint,bibnotes,amsmath,amssymb,aps,prb,showpacs,floatfix,superscriptaddress,bibliography]{revtex4-1}

\usepackage{amsmath,empheq}
\usepackage{amsfonts}
\usepackage{amssymb}
\usepackage{amsxtra}
\usepackage{mathtools}
\usepackage{xcolor}
\usepackage{graphicx}
\usepackage{subfigure}
\usepackage{dcolumn}
\usepackage{float}
\usepackage{bm}
\usepackage{hyperref} 
\hypersetup{breaklinks=true,colorlinks,citecolor=blue,linkcolor=blue,urlcolor=blue}

\newcommand{\nn}{\nonumber}

\newcommand{\bsigma}{\boldsymbol{\sigma}}
\DeclareMathAlphabet{\bi}{OML}{cmm}{b}{it}
\def\be{\begin{equation}}
\def\ee{\end{equation}}
\def\bearr{\begin{eqnarray}}
\def\eearr{\end{eqnarray}}

\def\bs{\boldsymbol}

\begin{document}
\title{Effect of chiral anomaly on the circular dichroism and Hall angle in doped and tilted Weyl semimetals}
\author{Ashutosh Singh}
\email{asingh.n19@gmail.com}
\affiliation{Department of Physics and astronomy, McMaster University, Hamilton, Ontario L8S 4M1, Canada}
\author{J. P. Carbotte}
\email{carbotte@mcmaster.ca}
\affiliation{Department of Physics and astronomy, McMaster University, Hamilton, Ontario L8S 4M1, Canada}
\affiliation{Canadian Institute for Advanced Research, Toronto, Ontario M5G 1Z8, Canada}

\date{\today}

\begin{abstract}
From the Kubo formula for transport in a tilted Weyl semimetal we calculate the absorptive part of the dynamic conductivity for both right and left handed circular polarized light. These depend on the real part of the longitudinal conductivity and the imaginary part of the transverse (Hall) conductivity.
We include the effect of the chiral anomaly which pumps charge from negative to positive chirality node when the usual ${\bs E}\cdot{\bs B}$ term is included in the electrodynamics and obtain analytic expressions. To calculate the Hall angle we further provide expressions for the imaginary part of longitudinal and real part of the transverse conductivity and compare results with and without the pumping term.
We also consider the case of a non centro symmetric Weyl semimetal in which the chiral nodes are displaced in energy by an amount $\pm \mathcal{Q}_0$. This leads to modification in dichroism and Hall angle which parallel the pumping case.
\end{abstract}

%
\maketitle
\section{Introduction}
\begin{figure}[ht!]
\includegraphics[width = \linewidth]{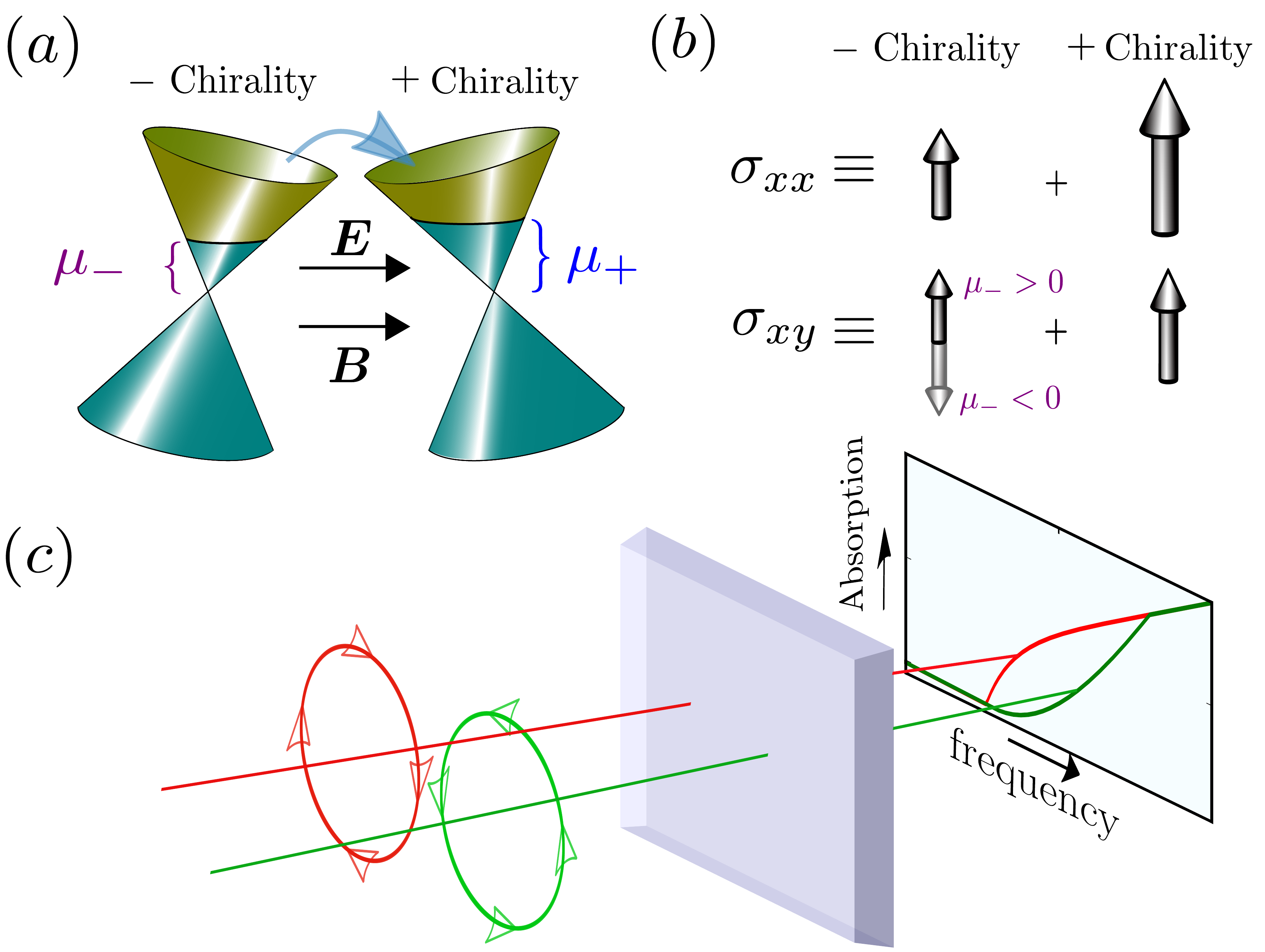} 
\caption{(a) Schematic of two tilted Weyl cones with nonzero chiral pumping. The positive chirality node and the negative chirality node are tilted anticlockwise and 
clockwise respectively and we include chiral pumping, both affect the Hall conductivity. 
Application of ${\bs E}\cdot{\bs B}$ term results in a net transfer of charge carriers from one node to the other (blue arrow), thereby modifying the chemical potential of the two nodes. The chemical potential of the negative and the positive chirality nodes are denoted as $\mu_-$ and $\mu_+$ respectively. 
(b) The total longitudinal $\left(\sigma_{xx}\right)$ and Hall $\left(\sigma_{xy}\right)$ conductivity is a sum of the contribution from
each chirality node. The upward and downward arrows refer to the positive and the negative contribution respectively. We consider a scenario where $\mu_+$ is always positive, whereas the sign of $\mu_-$ depends on the chiral pumping [Eq.~\eqref{mu_s}]. Therefore the contribution coming from positive chirality is always positive. The length of the arrows signifies the different doping levels in each of the two Weyl nodes. 
(c) The charge imbalance makes a non-vanishing contribution to longitudinal and Hall conductivity which leads to changes in the circular dichroism, i.e. the absorption of the right (red) and the left (green) circularly polarized light is modified by the chiral pumping.}
\label{fig1}
\end{figure}

In the last decade an area of research referred to as topological materials has seen phenomenal growth and has introduced 
many new ideas and concepts as well as materials with new functionalities \cite{Kane,Qi,Ashvin,*pickett,Vafek, bansil,burkov1}.
Of particular interest here are the Weyl semimetals \cite{Weng,Huang2015,PhysRevX.5.031013,Lv2015,Xu613,Yong,Soluyanov2015,PhysRevB.95.241108,PhysRevB.94.121113,Huang2016}. 
Low lying quasiparticles in these systems are well described by a relativistic energy dispersion introduced by Dirac with a characteristic Fermi velocity $v_F$, a material parameter which is usually much less than the speed of light.
The Dirac cones in these materials can be tilted with respect to the energy axis and there 
are two types. For type I ~(under tilted), the tilt $\mathcal{C}$ is less than the Fermi velocity, $v_F$ [See Fig.~\ref{fig1}(a)]. For $\mathcal{C}= v_F$
the cones have fully tipped and for $\mathcal{C}> v_F$ they are over tilted (type II)\cite{Soluyanov2015}. At charge neutrality the Fermi surface 
remains a single point at the node of the Dirac cone for type I while electron-hole pockets form at zero energy for type II.
Depending on the magnitude and tilt direction, one can observe important effects on the electromagnetic properties of Weyl semimetals. In particular, tilts modify
collective effects\cite{Detassis}, the anomalous Hall effect\cite{Mukherjee_2018,Zyuzin2016,Trescher}, the chiral magnetic effects\cite{Wurff}, the anomalous Nernst\cite{PhysRevB.96.115202,Saha2018} and they modify disorder effects\cite{PhysRevB.96.155121}.
Here we turn to the effect of tilt in type I Weyl semimetals on both the dichroism and to the Hall angle.

In the linear response model, the interband optical absorption background, ${\rm Re}\left[\sigma_{xx}\left(\Omega\right)\right]$\cite{Jules}, in a non-tilted Weyl semimetal, changes abruptly at $2\mu_0$, with $\mu_0$ being the chemical potential.
Below this point, the optical transitions contributing to the background are Pauli blocked, whereas, the interband absorption background recovers its undoped value for $\Omega \geq 2\mu_0$, and varies linearly with frequency afterwards\cite{Jules}. In type I Weyl, the jump in the absorption background at $\Omega = 2\mu_0$ is replaced by a quasi linear smooth rise in the interval $2\mu_0/(1+\mathcal{C}/v_F)<\Omega<2\mu_0/(1-\mathcal{C}/v_F)$. In this case, absorption is allowed at a lower frequency, $\Omega > 2\mu_0/(1+\mathcal{C}/v_F)$, as compared to non-tilted Weyl case and the undoped background is restored beyond $2\mu_0/(1-\mathcal{C}/v_F)$. In this same photon energy interval, with tilt and finite doping, the imaginary part of the dynamic transverse Hall conductivity $\left({\rm Im}\left[\sigma_{xy}\left(\Omega\right)\right]\right)$\cite{Mukherjee_im} becomes finite. 
This leads to dichroism in the absorption of circular polarized\cite{Mukherjee_circ} light, since right and left hand polarizations will have different 
conductivities $\left({\rm Re}\left[\sigma_{\pm}\left(\Omega\right)\right], ~\text{see}~ \text{Fig}.~\ref{fig1}(b)~\text{and}~ \text{Fig}.~\ref{fig1}(c)\right)$. 
One needs to remember that, while the contributions to ${\rm Im}\left[\sigma_{xy}\left(\Omega\right)\right]$ of both negative and positive chirality nodes have the same magnitude for a given $\mu_0$ and $\mathcal{C}$, they carry opposite 
sign~(Fig.\ref{fig1}(b)). Further, changing the relative sign of the tilt on one cone again changes the sign of ${\rm Im}\left[\sigma_{xy}\left(\Omega\right)\right]$ so that, if both are tilted clockwise or both anti-clockwise, the sum of ${\rm Im}\left[\sigma_{xy}\left(\Omega\right)\right]$ coming from the two cones will cancel if they have the same magnitude of $\mu_0$ and $\mathcal{C}$ and there will be no 
dichroism. To get a finite effect we need the cones to be oppositely tilted, one clockwise the other anticlockwise and this is the case we treat here.

An effect of particular importance, which is associated with Weyl semimetals, is the chiral anomaly\cite{Burkov_2015,Ashby,Zhou,Hosur,sid,Goswami,Hosur_Ashvin, Asteria2019}. The application of a parallel electric $(\bs E)$ and magnetic 
$(\bs B)$ field drives charge from a negative chirality node to a positive chirality node\cite{Ashby,Zhou,Hosur,PhysRevB.97.035403}. This leads to the two Weyl cones to effectively have different values of chemical potential and this alters their relative contribution to the dichroism when tilting is considered. 
This is illustrated in Fig.~\ref{fig1}(a). In the steady state, the chemical potential for the negative and the positive chirality nodes are denoted as $\mu_-$ and $\mu_+$  respectively. This leads to the emergence of new regions in the absorption spectra of ${\rm Re}\left[\sigma_{xx}\left(\Omega\right)\right]$ and ${\rm Im}\left[\sigma_{xy}\left(\Omega\right)\right]$, as there are four distinct critical frequencies at play: $2\mu_{-}/(1\pm\mathcal{C}/v_F)$ and $2\mu_{+}/(1\pm\mathcal{C}/v_F)$, corresponding to the two nodes. Therefore, by tuning the chiral pumping which in turn changes these frequencies, one expects to see interesting modifications in the absorption of left and right handed polarized light.
In this work, we ignore the possibility of chiral pumping resulting from the rotating electric field and the steady magnetic field\cite{Oka}, which is significant only when the electric field amplitude and the frequency are large.

This paper is organized as follows: In Sec.~\ref{Model}, we present the theoretical formulation including the discussion on system's Hamiltonian, node dependent chemical potential, and the formalism for calculating the interband optical conductivity. In Sec.~\ref{Weyl_cond} we calculate the optical conductivity for type I Weyl semimetals and provide analytical expressions for ${\rm Re}\left[\sigma_{xx}\left(\Omega\right)\right]$
and ${\rm Im}\left[\sigma_{xy}\left(\Omega\right)\right]$ and discuss the circular dichroism. 
The remaining parts, i.e. ${\rm Im}\left[\sigma_{xx}\left(\Omega\right)\right]$ and ${\rm Re}\left[\sigma_{xy}\left(\Omega\right)\right]$ are
calculated using Kramers-Kronig relations and their analytical expressions along with the discussion of the Hall angle in Sec.~\ref{Bulk_Hall}.
In Sec.~\ref{Q0}, we consider the case of broken inversion symmetry. In this case the negative chirality node is pushed up in energy by an amount $\mathcal{Q}_0$ while the positive chirality node is pushed down by the same amount.
This leads to an effective doping in the two cones describe by the chemical potential $\mu_{\pm} = \mu_0\pm\mathcal{Q}_0$. This is analogous to the pumping case with a different relationship between the two chemical potentials.
Finally in Sec.~\ref{summ} we summarize and draw conclusions.
\section{THEORETICAL MODEl}\label{Model}
\subsection{Hamiltonian}
The Hamiltonian for low-energy quasi-particles in a tilted Weyl semimetal with nodal index $s$ is given as,
\be\label{H_Weyl}
\hat {\mathcal H}^{s} =  \hbar\mathcal{C}_{s}k_z~\mathbb{I}_{2\times2} + s \hbar v_F{\bsigma}\cdot{\bs k}~,
\ee
where $\mathcal{C}_{s}$ denotes the tilt velocity, $\mathbb{I}_{2\times2}$ is the $2\times2$ unit matrix, $v_F$ is the Fermi velocity, $\bsigma$ consists of Pauli matrix triplets and ${\bs k}=(k_x,k_y,k_z)$, is the quasi-particle momentum. The energy eigenvalue is, therefore, given as, $\varepsilon^{\lambda}_{\bs k} = \hbar\mathcal{C}_{s}k_z + \hbar v_F\lambda|{\bs k}|$, with $\lambda = 1~(-1)$ for conduction~(valence) band.
The chirality $s$ is chosen to be positive~(negative) for anticlockwise~(clockwise) tilted node. The $|\mathcal{C}_{s}|<v_F$ case is characterized as a type-I Weyl semimetal whereas $|\mathcal{C}_{s}|>v_F$ refers to a type-II Weyl semimetal. In this work, we restrict ourselves to the former class of materials only, as the extension of these results to type-II semimetals is straightforward. In fact, much of the results to be presented in this work can readily be given for type II cases as well, following some recent works~(Ref.~[\onlinecite{Mukherjee_im}] and Ref.~[\onlinecite{Mukherjee_circ}], for example). This is of course allowed only when the charge imbalance between the two Weyl nodes has been properly addressed in the context of the type II semimetal, which is beyond the scope of the present work.

\subsection{Charge non-conservation: nodal chemical potential}
The chiral anomaly term, which is proportional to ${\bs E}\cdot{\bs B}$, pumps charge from negative chirality to the positive chirality node. In doing so, the density of charge in the region of Weyl nodes with chirality $s~(\equiv\pm)$ is changed by\cite{Ashby,Zhou,Hosur},
\be
\Delta n_{s} = s \frac{e^2}{4\pi^2\hbar^2}\left({\bs E}\cdot{\bs B}\right)\tau_{\nu},
\ee
where $\tau_{\nu}$ is the inter-nodal relaxation time and is assumed large compared with the intra-node scattering
rate so that thermal equilibrium is reached inside each nodal cone.
To start with, we consider the case of tilted Weyl semimetal when the chiral term is not present, such that the total charge carrier is given by,
\be\label{n_0}
n_0 = \frac{1}{6\pi^2}\frac{1}{\hbar^3 v_F^3 }\frac{\mu^3_0}{\left[\left(\mathcal{C}^2_{s}/v_F^2\right) - 1\right]^2},
\ee
where $\mu_0$ is the chemical potential without the chiral term which is taken to be positive for definitiveness. This is modified when the chiral pumping is included and the total number of charge carriers in node $s$ is given by,
\be\label{n_s}
n_{s} = n_0 + s \frac{e^2}{4\pi^2\hbar^2}\left({\bs E}\cdot{\bs B}\right)\tau_{\nu}.
\ee
Now using Eq.~\eqref{n_0} and Eq.~\eqref{n_s}, we find the chiral chemical potential to be,
\be\label{mu_s}
\mu_{s} = \left(\mu^3_0 + s \mu_p^3\right)^{1/3},
\ee
where, $\mu^3_p = 3e^2\hbar v_F^3\left({\bs E}\cdot{\bs B}\right)\tau_{\nu}\left[\left(\mathcal{C}^2_{s}/v_F^2\right) - 1\right]^2/2$ and is assumed to be 
positive. The charge transfer between the two nodes is illustrated in Fig.~\ref{fig1}(a). For negligible chiral pumping,  $\mu_p\to 0$, which implies $\mu_{\pm}\approx \mu_0$.
In the negative chirality node, the effective chemical potential becomes negative for $\mu_p > \mu_0$ and this node is hole doped. In order to proceed further in this case, we simply replace $\mu_-\to -|\mu_-|$.
The magnitude of $\mu_p$ for non tilted Weyl nodes have been estimated theoretically in Ref.~[\onlinecite{Ashby}] and found to be equal to $9.18$ meV. Inclusion of tilt leads to multiplication of a factor $\left[\left(\mathcal{C}^2_{s}/v_F^2\right) - 1\right]^2$ which results in $\mu_p\approx 7.5$ meV for $\mathcal{C}/v_F = 1/2$. For simplicity, we set $v_F$ and $\hbar$ to be equal to 1, such that $\mathcal{C}/v_F\to\mathcal{C}$, which serves as the tilt parameter.

\subsection{Optical conductivity}
The chiral optical conductivity can be given as \cite{mahan2010many}, $
\sigma^s_{\alpha\beta} = i\Pi^s_{\alpha\beta}\left(i\Omega_m\right)/{\Omega}~,
$
where the finite temperature ($T$) current-current correlation function is defined as,
\be
\Pi^s_{\alpha\beta} = \sum_{\omega_n}\int \frac{d{\bs k}}{\beta}{\rm Tr}\left[\hat J^s_{{\bs k}\alpha}\hat G^s_{\bs k}\left(i\omega_n+i\Omega_m\right)\hat J^s_{{\bs k}\beta}\hat G^s_{\bs k}\left(i\omega_n\right)\right],
\ee
with $i\Omega_m\to \Omega + i0^+$, $\beta = \left(k_BT\right)^{-1}$, $k_B$ being the Boltzmann constant, $d{\bs k}=d^3k/\left(2\pi\right)^3$, the current operator, $\hat{\bs J}^s_{\bs k} = \nabla_{\bs k}\hat{\mathcal H}^s$
and the Green's function is expressed as, $\hat G\left({\bs k},z\right) = \left(z~\mathbb{I}_{2\times2} - \hat {\mathcal{H}}\right)^{-1}$, which for the Hamiltonian described in Eq.~\eqref{H_Weyl}, can be expressed explicitly as,
\be\label{green}
\hat G^s_{\bs k}\left(z\right) = \frac{1}{\Delta}\begin{pmatrix}
                                       z + \left(s-\mathcal{C}_s\right)k_z & s\left(k_x-ik_y\right)   \\
									s\left(k_x + ik_y\right) & z-\left(s+\mathcal{C}_s\right)k_z
                                        \end{pmatrix},
\ee
where we have $\Delta = \left(z-\mathcal{C}_sk_z\right)^2-|{\bs k}|^2$.
The spectral function can be extracted using the following relation,
\be\label{green_spectral}
\hat G^s_{\bs k}\left(z\right) = \int_{-\infty}^{\infty} \frac{d\omega}{2\pi}\frac{\hat A^s_{\bs k}\left(\omega \right)}{z-\omega}.
\ee
Using, Eq.~\eqref{green} and Eq.~\eqref{green_spectral}, we see that,
\bearr\label{spectral}\nn
\hat A^s_{\bs k}\left(\omega \right) &=& \frac{\pi}{|{\bs k}|}\begin{pmatrix}
                                       \omega + \left(s-\mathcal{C}_s\right)k_z & s\left(k_x-ik_y\right)   \\
									s\left(k_x + ik_y\right) & \omega-\left(s+\mathcal{C}_s\right)k_z
                                        \end{pmatrix}\\
&\times&\left[\delta\left(\omega-\mathcal{C}_sk_z-|{\bs k}|\right)-\delta\left(\omega-\mathcal{C}_sk_z+|{\bs k}|\right)\right]~.
\eearr
Using the above expression and that $\hat J_{{\bs k}x} = e\hat\sigma_{x}$, we have,
\bearr\nn
\Pi^s_{xx}\left(i\Omega_m\right) &=& e^2\int d{\bs k}\int\frac{d\omega}{2\pi}\int\frac{d\omega^{\prime}}{2\pi}~\frac{f\left(\omega^{\prime}\right)-f\left(\omega\right)}{i\Omega_m-\omega+\omega^{\prime}}\\
&\times&
{\rm Tr}\left[\hat \sigma_x\hat A^s_{\bs k}\left(\omega\right)\hat \sigma_x\hat A^s_{\bs k}\left(\omega^{\prime}\right)\right]~,
\eearr
where, $f(x) = \left(1 + e^{\beta\left(x-\mu_s\right)}\right)^{-1}$, provided the chirality dependent chemical potential, $\mu_s$, is finite. 

In this work, we are mainly interested in the interband optical conductivity.
Therefore, focusing only on the interband contribution, the real part of the total conductivity can be simplified and is given as\cite{Jules},
\bearr\nn\label{rexx}
\frac{{\rm Re}[\sigma_{xx}(\Omega)]}{e^2\mu_0} &=& \frac{\Omega}{64\pi\mu_0}\sum_{s=\pm}\int_{-1}^{1}dx\left(1+x^2\right)\\\nn
&\times&\left(f\left[- \frac{\Omega}{2}\left(1-\mathcal{C}_sx\right)\right]-f\left[\frac{\Omega}{2}\left(1+\mathcal{C}_sx\right)\right]\right),\\
\eearr
where we have divided both sides of the equation with $\mu_0$, the chemical potential in the absence of the chiral pumping.
In a very similar manner we can write down the imaginary part of the Hall conductivity\cite{Mukherjee_im}, i.e.,
\bearr\nn\label{imxy}
\frac{{\rm Im}[\sigma_{xy}(\Omega)]}{e^2\mu_0} &=& -\frac{\Omega}{32\pi\mu_0}\sum_{s=\pm}s\int_{-1}^{1}dx~x\\\nn
&\times&\left(f\left[- \frac{\Omega}{2}\left(1-\mathcal{C}_sx\right)\right]-f\left[\frac{\Omega}{2}\left(1+\mathcal{C}_sx\right)\right]\right).\\
\eearr
\section{Optical conductivity: Tilted Weyl Semimetal}\label{Weyl_cond}
\begin{figure*}[ht]
\includegraphics[width = 0.9\linewidth]{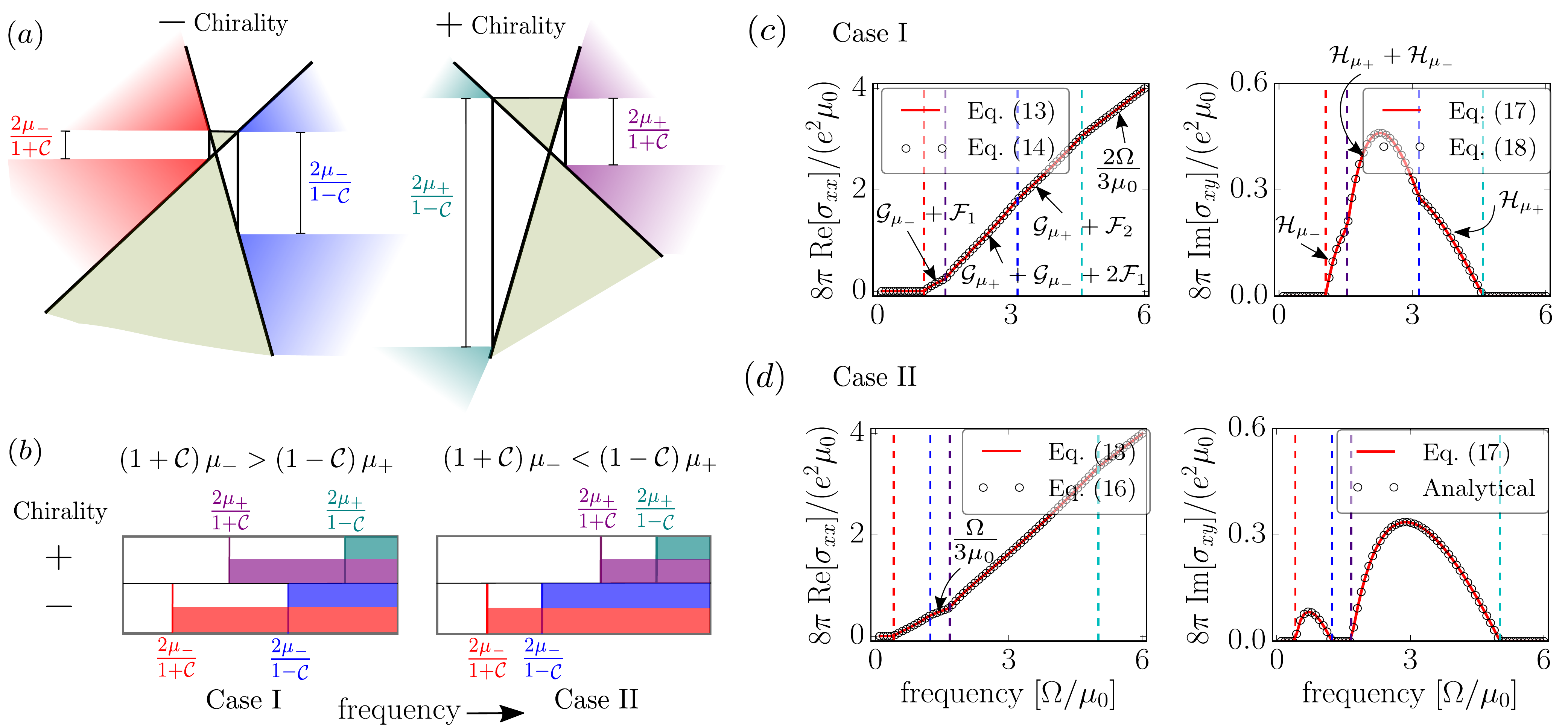} 
\caption{(a) Schematic for various transitions involved in the presence of charge imbalance. (b) Contributions coming from different frequency intervals in the positive and negative chirality nodes for two cases, namely $\left(1+\mathcal{C}\right)\mu_- > \left(1-\mathcal{C}\right)\mu_+$ [case I] and $\left(1+\mathcal{C}\right)\mu_- < \left(1-\mathcal{C}\right)\mu_+$ [case II] have been illustrated with distinct colors shading. For example, below the frequency value $\Omega^{\mu_-}_L \left(=2\mu_-/{\left(\mathcal{C} + 1\right)}\right)$, we do not expect the interband optical conductivity to be finite due to complete Pauli blocking. In case I, each node contributes partially in the frequency regime $2\mu_-/{\left(\mathcal{C} + 1\right)}< \Omega< 2\mu_-/{\left(1-\mathcal{C}\right)}$. For frequencies greater than $2\mu_+/{\left(1-\mathcal{C}\right)}$, Pauli blocking is completely removed and we expect full contribution from both the nodes. This is indeed the case as can be seen in (c),(d) where we have shown the real part of the longitudinal conductivity and the imaginary part of Hall conductivity as a function of frequency for $\mu_p = 0.8\mu_0$, which falls under case I and $\mu_p = 0.99\mu_0$ for the other case, respectively. Here we have considered $\mathcal{C_-}\left( = -\mathcal{C_+}\right) = \mathcal{C} = 0.5$ and $\mu_0$ is chosen as the normalization parameter.}
\label{fig2}
\end{figure*}

In this section we apply the formalism discussed in the previous section to calculate the longitudinal and Hall conductivities in a tilted Weyl semimetal.
Using Eq.~\eqref{rexx}, the real part of dynamic longitudinal conductivity at $T = 0$ for chiral node $s$ is given as,
\bearr\nn\label{xx}
\frac{{\rm Re}[\sigma^s_{xx}(\Omega)]}{e^2\mu_0/(8\pi)} &=& \frac{\Omega}{8\mu_0}\int_{-1}^{1}dx\left(1+x^2\right)
\bigg(\Theta\left[\mu_s + \frac{\Omega}{2}\left(1-\mathcal{C}_sx\right)\right]\\
&-&\Theta\left[\mu_s -\frac{\Omega}{2}\left(1+\mathcal{C}_sx\right)\right]\bigg),
\eearr
and the total longitudinal optical conductivity is given by,  ${\rm Re}[\sigma_{xx}(\Omega)] = {\rm Re}[\sigma^+_{xx}(\Omega)] + {\rm Re}[\sigma^-_{xx}(\Omega)]$. For definitiveness, we have considered $\mathcal{C_-}\left( = -\mathcal{C_+}\right) = \mathcal{C}, $ which is generally true for real materials when the inversion symmetry is preserved\cite{Wurff,PhysRevB.96.115202}.
%
%
The integral in Eq.~\eqref{xx} can be solved analytically, with the following defining cases: 1) Case I, where $(1 + \mathcal{C})\mu_- > (1 - \mathcal{C})\mu_+$~, and 2) Case II where~$(1 + \mathcal{C})\mu_- < (1 - \mathcal{C})\mu_+$~. These two cases have been illustrated in Fig.~\ref{fig2}(b), with contribution coming from each chirality node in different frequency intervals shown explicitly.
%
%
%
%
%
%
For Case I, we have,
\begin{empheq}[left={\dfrac{{\rm Re}[\sigma_{xx}(\Omega)]}{e^2\mu_0/(8\pi)}=\empheqlbrace}]{alignat=4}\nn
0, & \quad & \Omega < \Omega^{\mu_-}_L\\\nn
\mathcal{G}_{\mu_-} + \mathcal{F}_1,& \quad &\Omega^{\mu_-}_L < \Omega < \Omega^{\mu_+}_L\\\nn
\sum_{s=\pm}\mathcal{G}_{\mu_s} + 2\mathcal{F}_1, & \quad &\Omega^{\mu_+}_L < \Omega < \Omega^{\mu_-}_U\\\nn
\mathcal{G}_{\mu_+} + \mathcal{F}_2, & \quad & \Omega^{\mu_-}_U < \Omega < \Omega^{\mu_+}_U\\
2\Omega/\left(3\mu_0\right), & \quad & \Omega >  \Omega^{\mu_+}_U \label{Re_xx1}
\end{empheq}
where we have defined, 
\be
\mathcal{G}_{\mu_s} = \frac{\mu^2_s\left(3\Omega-2\mu_s\right)}{6\mu_0\mathcal{C}^3\Omega^2}-\frac{\left(1 + \mathcal{C}^2\right)\mu_s}{4\mu_0\mathcal{C}^3},
\ee
$\mathcal{F}_1 = \left[1+\mathcal{C}^2\left(3+4~\mathcal{C}\right)\right]\Omega/\left(24\mu_0\mathcal{C}^3\right)$, $\mathcal{F}_2 = \left[1+3~\mathcal{C}^2\left(1+4~\mathcal{C}\right)\right]\Omega/\left(24\mu_0\mathcal{C}^3\right)$.
For notational simplicity, we define the transitions frequencies, $\Omega^{\mu_{\pm}}_{L} = 2\mu_{\pm}/\left(1+\mathcal{C}\right)$ and $\Omega^{\mu_{\pm}}_{U} = 2\mu_{\pm}/\left(1-\mathcal{C}\right)$~$\left(\text{see Fig.~\ref{fig2}(a)}\right)$.
\\
For Case II, we have,
\begin{empheq}[left={\dfrac{{\rm Re}[\sigma_{xx}(\Omega)]}{e^2\mu_0/(8\pi)}=\empheqlbrace}]{alignat=4}\nn
0, ~~& \quad & \Omega < \Omega^{\mu_-}_L\\\nn
\mathcal{G}_{\mu_-} + \mathcal{F}_1,~~& \quad &\Omega^{\mu_-}_L < \Omega < \Omega^{\mu_-}_U\\\nn
\Omega/\left(3\mu_0\right),~~& \quad &\Omega^{\mu_-}_U < \Omega < \Omega^{\mu_+}_L\\\nn
\mathcal{G}_{\mu_+} + \mathcal{F}_2,~~& \quad & \Omega^{\mu_+}_L < \Omega < \Omega^{\mu_+}_U\\
2\Omega/\left(3\mu_0\right),~~& \quad & \Omega >  \Omega^{\mu_+}_U \label{Re_xx2}
\end{empheq}
%
\\
The imaginary part of Hall conductivity in $T\to 0$ limit is provided by Eq.~\eqref{imxy}, which is given as,
\bearr\nn
\frac{{\rm Im}[\sigma_{xy}(\Omega)]}{e^2/(8\pi)} &=& -\frac{\Omega}{4}\sum_{s=\pm}s\int_{-1}^{1}dx~x\Bigg(\Theta\left[\mu_s + \frac{\Omega}{2}\left(1-\mathcal{C}_sx\right)\right]\\
&-&\Theta\left[\mu_s -\frac{\Omega}{2}\left(1+\mathcal{C}_sx\right)\right]\Bigg)~.
\eearr
%
%
%
%
%
%
Again the above integral can easily be carried out and we have the following analytical expression for Case I,
\begin{empheq}[left={\dfrac{{\rm Im}[\sigma_{xy}(\Omega)]}{e^2\mu_0/(8\pi)}=\empheqlbrace}]{alignat=4}\nn
0, & \quad & \Omega < \Omega^{\mu_-}_L\\\nn
\mathcal{H}_{\mu_-} ,& \quad &\Omega^{\mu_-}_L < \Omega < \Omega^{\mu_+}_L\\\nn
\sum_{s=\pm}\mathcal{H}_{\mu_s},& \quad &\Omega^{\mu_+}_L < \Omega < \Omega^{\mu_-}_U\\\nn
\mathcal{H}_{\mu_+},& \quad & \Omega^{\mu_-}_U < \Omega < \Omega^{\mu_+}_U\\\label{Im_xy1}
0, & \quad & \Omega > \Omega^{\mu_+}_U  
\end{empheq}

where, 
\be
\mathcal{H}_{\mu_s} = \frac{\mu_s}{2\mu_0}\frac{\left(\Omega - \mu_s\right)}{\Omega~\mathcal{C}^2} + \frac{\left(\mathcal{C}^2-1\right)\Omega}{8\mu_0\mathcal{C}^2}.
\ee
For case II, Eq.~\eqref{Im_xy1} is modified in such a way that we have $\mathcal{H}_{\mu_-}$ in the frequency regime $\Omega^{\mu_-}_L<\Omega<\Omega^{\mu_-}_U$ and $\mathcal{H}_{\mu_+}$ in the frequency regime $\Omega^{\mu_+}_L<\Omega<\Omega^{\mu_+}_U$.
These are the only two interval where the ${\rm Im}[\sigma_{xy}(\Omega)]$ is non-zero in this particular case.

For special case when we have $(1 + \mathcal{C})\mu_- = (1 - \mathcal{C})\mu_+$, the expression for the longitudinal and the Hall conductivity is still given
by Eq.~\eqref{Re_xx1} and Eq.~\eqref{Im_xy1} respectively, except that the condition $\Omega^{\mu_+}_L < \Omega < \Omega^{\mu_-}_U$ now becomes redundant as in this particular case we have,
$\Omega^{\mu_+}_L = \Omega^{\mu_-}_U$.
Another case that could be realized in practice is when we have $\mu_p > \mu_0$.
In this case, for ${\rm Re}[\sigma_{xx}(\Omega)]$, the expressions provided in Eq.~\eqref{Re_xx1} and Eq.~\eqref{Re_xx2} can be used with only a replacement, $\mu_-\to\mu^{\prime}_-$, where  $\mu^{\prime}_- = \left(\mu^3_p - \mu^3_0\right)^{1/3}$. 
For the Hall conductivity, we need to replace $\mathcal{H}_{\mu_-}$ with $-\mathcal{H}_{\mu^{\prime}_-}$ in both cases, i.e. Case I [Eq.~\eqref{Im_xy1}] as well as Case II.

In Fig.~\ref{fig2} we show explicit results for two values of $\mu_p$ namely $\mu_p = 0.8\mu_0$ in frames (c) and $\mu_p = 0.99\mu_0$ in frames (d). The left frames in (c) and (d) are for the real part of the dynamic longitudinal conductivity while the right are for imaginary part of the dynamic Hall conductivity. Frame (a) shows a schematic for the optical transitions possible in the presence of charge imbalance. In frame (b), two possible cases have been shown, clearly demonstrating the contribution from each node separately in different frequency intervals.  Returning to frames (c) and (d), the four dashed vertical lines identify the critical frequencies $\Omega^{\mu_{\pm}}_L = 2\mu_{\pm}/(1+\mathcal{C})$ and
$\Omega^{\mu_{\pm}}_U = 2\mu_{\pm}/(1-\mathcal{C})$. With $\mathcal{C} = 1/2$ for definiteness, $\Omega^{\mu_{-}}_L$ is red, $\Omega^{\mu_{-}}_U$ is blue, $\Omega^{\mu_{+}}_L$ is indigo and $\Omega^{\mu_{+}}_U$ is cyan. In frame (b) $\left(1 + \mathcal{C}\right)\mu_{-}>\left(1 - \mathcal{C}\right)\mu_{+}$, with $\mu_p = 0.8\mu_0$ neither chirality node contributes to the longitudinal conductivity (left frame) below $\Omega^{\mu_{-}}_L$ and in the next interval $\Omega^{\mu_{-}}_L$ to $\Omega^{\mu_{+}}_L$, only the negative chirality node is involved. In the third and fourth intervals both nodes enter and in the fifth above $\Omega^{\mu_{+}}_U$ we get back the no tilt, no doping interband background of the longitudinal conductivity.

For the transverse Hall conductivity shown in the right frame the situation is similar but now the contribution from the negative chirality node ends at $\Omega^{\mu_{-}}_U$ (blue vertical line). Above this energy in the fourth interval the positive chirality node continues till $\Omega^{\mu_{+}}_U$ (cyan vertical line) above which the Hall conductivity vanishes. For no pumping $\mu_+ = \mu_- = \mu_0$ and red and indigo vertical dashed lines merge as do blue and cyan and both chirality nodes contribute equally in the energy interval where they do not vanish. By contrast pumping has created two new energy interval in which only one of the two nodes contribute to the Hall conductivity.  This will have a direct effect on the dichroism for absorption of circular polarized light as we will see later.

In frame (d) of Fig.~\ref{fig2} we present as additional set of results for the case $\left(1 + \mathcal{C}\right)\mu_{-}<\left(1 - \mathcal{C}\right)\mu_{+}$ ($\mu_p = 0.99\mu_0$), for which Eq.~\eqref{Re_xx2} applies rather than Eq.~\eqref{Re_xx1}. Continuing with our discussion of the Hall conductivity (right frame), the non zero contribution to ${\rm Im}[\sigma_{xy}\left(\Omega\right)]$ in the second energy interval between red and blue vertical lines is entirely due to the negative chirality node which is followed by a third region where ${\rm Im}[\sigma_{xy}\left(\Omega\right)]$ vanishes. Beyond this, in the fourth energy interval only the positive chirality node contributes and finally above this, in a fifth interval there is no Hall conductivity. Thus, in the case for $\mu_p = 0.99\mu_0$ we have managed to separate out the contributions of positive chirality node and negative chirality node to distinct regions of frequencies. In reference to the longitudinal conductivity (left frame) we note that in the third energy region only the negative chirality node is involved and we are above $\Omega^{\mu_{-}}_U$ so that the no doping no tilting interband background applies and this straight line is $\Omega/{3\mu_0}$, exactly half the value which applies above the last cyan vertical line where the two nodes are involved.

Till now we have discussed only the interband conductivity. In the low frequency regime, the intraband contribution dominates which, in the clean limit, is expressed as\cite{Jules}
$: {\rm Re}[\sigma^{\rm intra}_{xx}\left(\Omega\right)] = \mathcal{D}\delta\left(\Omega\right)$, where, the Drude weight arising from a node
of chirality $s$ is given as,
\be\label{intra}
\mathcal{D}^s = \frac{e^2\mu^2_s}{8\pi\mathcal{C}^3}\left(\frac{2~\mathcal{C}}{1-\mathcal{C}^2}-\log\left[\frac{1+\mathcal{C}}{1-\mathcal{C}}\right]\right),
\ee
such that the total Drude weight is given as, $\mathcal{D} = \mathcal{D}^+ + \mathcal{D}^-$. There is no such intraband contribution to the Hall conductivity.
\\

{ \it  Dichroism:-} Right and left circularly polarized light gets absorbed by different amounts due to the presence of transverse optical conductivity. If the absorptive part of the optical conductivities are denoted as, $\sigma_+$ and $\sigma_-$ respectively, we define,
\be\label{dichroism}
\sigma_{\pm}\left(\Omega\right) = {\rm Re}\left[\sigma_{xx}\left(\Omega\right)\right] \mp {\rm Im}\left[\sigma_{xy}\left(\Omega\right)\right]~,
\ee
hence one can obtain the analytical expression by simply adding/subtracting the ${\rm Re}\left[\sigma_{xx}\left(\Omega\right)\right]$ and ${\rm Im}\left[\sigma_{xy}\left(\Omega\right)\right]$ parts.
\begin{figure}[ht]
\includegraphics[width = \linewidth]{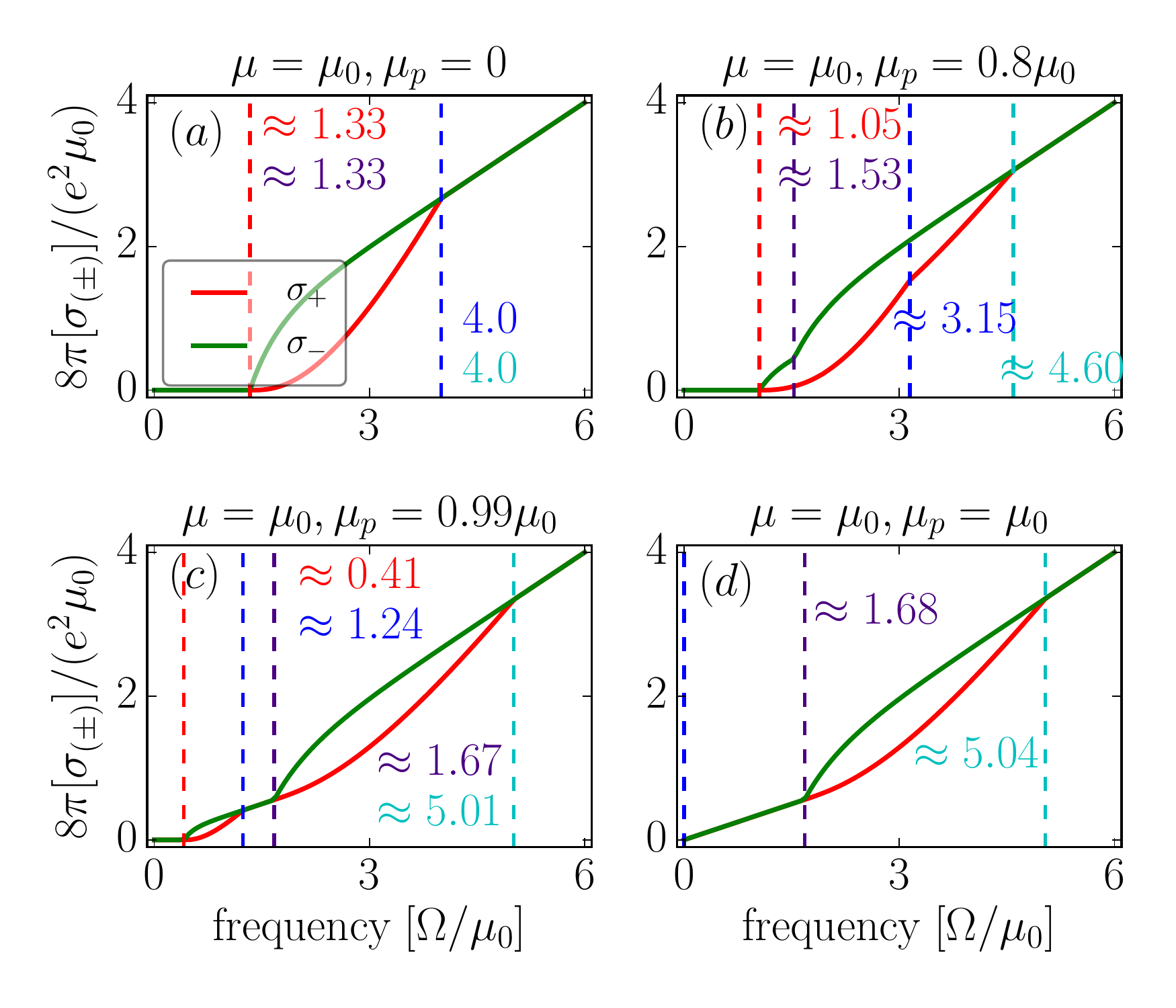} 
\caption{Optical conductivity, ($\sigma_{\pm}$) as seen by left and right circularly polarized light as a function of frequency for (a) $\mu_p = 0$, (b) $\mu_p = 0.8\mu_0$, (c) $\mu_p = 0.99\mu_0$, and (d) $\mu_p = \mu_0$. For $\mu_p\to \mu_0$, $\mu_-\to 0$, implying that for the negative node optical transitions start at zero frequency.  Increasing frequency beyond $\Omega^{\mu_+}_U$, we see that both the
nodes contribute, as expected. Color coded numbers identify the values of the critical frequencies defining the various photon ranges. Other parameters are the same as those of Fig.~\ref{fig2}.}
\label{fig3}
\end{figure}

In Fig.~\ref{fig3} we present results for the absorptive part of the dynamic conductivity associated with right and left handed polarized light. 
For vanishing chiral pumping the chemical potential in each Weyl node is the same and each node contributes equally to both longitudinal and transverse conductivity. There is no absorption below photon energy of $4\mu_0/3$ $\left(\text{frame (a)}\right)$. 
This is followed with a region of dichroism. Above photon energy of $4\mu_0$ there remains no dichroism  and the conductivity is due entirely to the longitudinal part, as the Hall contribution is zero. In the units used, this contribution is linear with slope $2/3$. 
When $\mu_p$ is increased to $0.8\mu_0$, shown in frame $(b)$, the region of no absorption below the first vertical line at energy $4\mu_-/3$ is reduced as compared with the no pumping case. 
It is followed by a region where only the negative chirality node contributes to the absorption and the dichroism is due only to this node. At the second vertical line the positive chirality node begins to contribute and both nodes contribute to the dichroism. At the third vertical line the negative chirality node no longer has a finite off diagonal (transverse) part and the dichroism is due entirely to the positive chirality node. Above the fourth vertical dashed line the system returns to its no doping no tilting value as in frame $(a)$.

In frame $(c)$ the pumping has been increased to $0.99\mu_0$. The small photon energy region of no absorption has again been reduced. It is followed by a region of dichroism extending between the first two vertical dashed lines. Here only the negative chirality node contributes. This is followed by a third region where there is no dichroism and the conductivity is linear in photon energy with slope $1/3$ as again only the negative chirality node is involved. In the region above the third vertical line both nodes are now involved but the dichroism comes only from the positive chirality node. For the final region above the fourth vertical line there is no dichroism at all since the transverse Hall conductivity is zero for both nodes. Finally we point out that the region between the first two vertical lines behaves exactly as in frame $(a)$ but now only the negative chirality node is involved. For the last frame the doping in the negative chirality node is zero and the absorptive Hall conductivity for this node is zero at all photon energies. This contribution now starts at zero energy and the conductivity shows no dichroism till the second vertical dashed line is reached and the positive chirality node also contributes. There is dichroism in this region due only to the positive chirality node until the third vertical dashed line is reached at which point the conductivity returns to its no doping, no tilting value. 

%

Pumping has made important changes to both longitudinal and transverse (Hall) dynamic conductivity. While for $\mu_p = 0$ both Weyl nodes contribute equally to both these quantities, this is no longer the case when $\mu_p$ is non zero. Finite $\mu_p$ can separate out regions of photon energies in which a single chirality node contributes and others where both contribute but not equally. In particular the dichroism can be entirely due to one node or the other or a combination of both but not in equal proportion.
In all cases there are kinks in the conductivities as we move from one region of photon energy to another. 

\section{Bulk Hall angle}\label{Bulk_Hall}
\begin{figure}[ht]
\includegraphics[width = \linewidth]{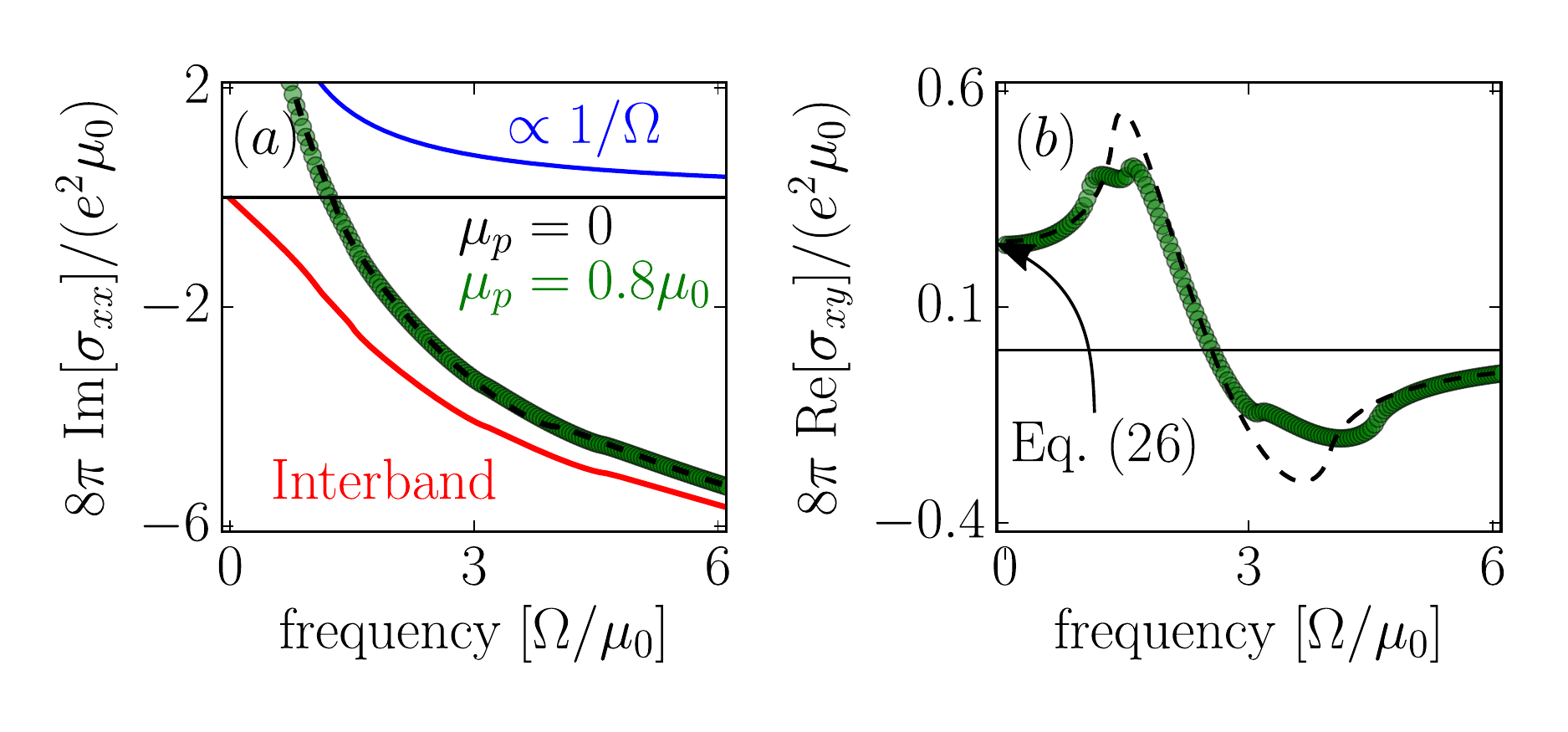} 
\caption{(a) The imaginary part of longitudinal and (b) real part of the Hall conductivity is shown as a function of frequency. The bubbled line in (a) refers to the total sum of the contribution from the interband (red) and the intraband (blue) conductivity, black dashed line is for $\mu_p = 0$ and green circles for $\mu_p = 0.8\mu_0$. In (b), the dc value of Hall conductivity is consistent with Eq.~\eqref{xy_dc}. Other parameters are the same as those of Fig. \ref{fig2}.}
\label{fig4}
\end{figure}

Having discussed the absorption of circularly polarized light, we now proceed to study the bulk Hall angle.
The dynamic bulk Hall angle can be defined as\cite{PhysRevB.73.245411,PhysRevB.75.165407,PhysRevLett.78.1572},
\be\label{Hall}
\Theta_H\left(\Omega\right) = \tan^{-1}\left({\rm Re}\left[\frac{\sigma_{xy}\left(\Omega\right)}{\sigma_{xx}\left(\Omega\right)}\right]\right),
\ee
where $\sigma_{\alpha\beta}\left(\Omega\right) = {\rm Re}\left[\sigma_{\alpha\beta}\left(\Omega\right)\right]+i{\rm Im}\left[\sigma_{\alpha\beta}\left(\Omega\right)\right]$, 
such that $\left(\alpha,\beta\right)\in\left\lbrace x,y\right\rbrace$.
Apart from the real part of longitudinal and imaginary part of Hall conductivity discussed in the previous sections, we also need the imaginary part of longitudinal and real part of Hall conductivity for the calculation of the Hall angle.
In order to calculate these quantities we rely on the Kramers-Kronig relation for conductivities which are given as,
\bearr\nn\label{KK}
{\rm Im}\left[\sigma_{xx}\left(\Omega\right)\right] &=& -\frac{2\Omega}{\pi}\mathcal{P}\int_0^{\Lambda}d\Omega^{\prime}\frac{{\rm Re}\left[\sigma_{xx}\left(\Omega^{\prime}\right)\right]}{{\Omega^{\prime}}^2-\Omega^2},\\
{\rm Re}\left[\sigma_{xy}\left(\Omega\right)\right] &=& \frac{2}{\pi}\mathcal{P}\int_0^{\Lambda}d\Omega^{\prime}\frac{\Omega^{\prime}{\rm Im}\left[\sigma_{xy}\left(\Omega^{\prime}\right)\right]}{{\Omega^{\prime}}^2-\Omega^2}~,
\eearr
where $\mathcal{P}$ refers to the principle value of the integral and $\Lambda~\left(\gg \Omega\right)$ is the ultra-violet cutoff. For the case $\left(1+\mathcal{C}\right)\mu_- > \left(1-\mathcal{C}\right)\mu_+$, substituting Eq. \eqref {Re_xx1} and Eq. \eqref {Im_xy1} in Eq. \eqref{KK}, we find the corresponding Kramers-Kronig counterparts as given below,
\begin{widetext}
\bearr\nn
\frac{{\rm Im}[\sigma_{xx}\left(\Omega\right)]}{e^2\mu_0/\left(8\pi\right)} &=& -\frac{\Omega}{24\mu_0\pi \mathcal{C}^3}\sum_{s =\pm}\Bigg[ \left(\frac{8\mu^3_s}{\Omega^3}+ \frac{6\left(1+\mathcal{C}^2\right)\mu_s}{\Omega}\right)\log\left[\frac{\left(\Omega^{\mu_s}_L-\Omega\right)\left(\Omega^{\mu_s}_U+\Omega\right)}{\left(\Omega^{\mu_s}_L+\Omega\right)\left(\Omega^{\mu_s}_U-\Omega\right)}\right] 
+ \frac{12\mu^2_s}{\Omega^2}\log\left[\frac{{\Omega^{\mu_s}_L}^2}{{\Omega^{\mu_s}_U}^2}\frac{\left({\Omega^{\mu_s}_U}^2-\Omega^2\right)}{\left({\Omega^{\mu_s}_L}^2-\Omega^2\right)}\right]\\\nn
&+&\frac{16\mu^3_s}{\Omega^2}\left(\frac{1}{\Omega^{\mu_s}_L}-\frac{1}{\Omega^{\mu_s}_U}\right)\Bigg] -\frac{1}{\pi\mu_0}\left[\mathcal{F}_1\left(2\log\left[\frac{{\Omega^{\mu_-}_U}^2-\Omega^2}{{\Omega^{\mu_+}_L}^2-\Omega^2}\right]+\log\left[\frac{{\Omega^{\mu_+}_L}^2-\Omega^2}{{\Omega^{\mu_-}_L}^2-\Omega^2}\right]\right)+\mathcal{F}_2\log\left[\frac{{\Omega^{\mu_+}_U}^2-\Omega^2}{{\Omega^{\mu_-}_U}^2-\Omega^2}\right]\right]\\
& -&\frac{2\Omega}{3\mu_0\pi}\log\left[\frac{{\Lambda}^2-\Omega^2}{{\Omega^{\mu_+}_U}^2-\Omega^2}\right] + \frac{16\mathcal{D}}{e^2\mu_0\Omega}~,
\eearr
and,
\bearr\nn
\frac{{\rm Re}[\sigma_{xy}\left(\Omega\right)]}{e^2\mu_0/\left(8\pi\right)}  &=& \frac{\left(\mathcal{C}^2-1\right)}{4\mu_0\pi \mathcal{C}^2}\left(\frac{\Omega}{2}\log\left[\frac{\left(\Omega^{\mu_-}_L+\Omega\right)\left(\Omega^{\mu_-}_U-\Omega\right)\left(\Omega^{\mu_+}_L+\Omega\right)\left(\Omega^{\mu_+}_U-\Omega\right)}{\left(\Omega^{\mu_-}_L-\Omega\right)\left(\Omega^{\mu_-}_U+\Omega\right)\left(\Omega^{\mu_+}_L-\Omega\right)\left(\Omega^{\mu_+}_U+\Omega\right)}\right] + \Omega^{\mu_+}_U+\Omega^{\mu_-}_U-\Omega^{\mu_+}_L-\Omega^{\mu_-}_L\right) \\
&+& \sum_{s=\pm}\frac{\mu_s}{2\mu_0\pi \mathcal{C}^2}\log\left[\frac{{\Omega^{\mu_s}_U}^2-\Omega^2}{{\Omega^{\mu_s}_L}^2-\Omega^2}\right] - \frac{\mu^2_s}{2\mu_0\pi \mathcal{C}^2\Omega}\log\left[\frac{\left(\Omega^{\mu_s}_U-\Omega\right)\left(\Omega^{\mu_s}_L+\Omega\right)}{\left(\Omega^{\mu_s}_U+\Omega\right)\left(\Omega^{\mu_s}_L-\Omega\right)}\right]~.
\eearr
\end{widetext}
Similarly, we can obtain the Kramers-Kronig pairs for other cases.
In the DC limit (photon energy $\Omega = 0$) the expression for ${\rm Re}\left[\sigma_{xy}\left(\Omega\right)\right]$ greatly simplifies and reduces to,
\be\label{xy_dc}
\frac{{\rm Re}[\sigma_{xy}\left(0\right)]}{e^2\mu_0/\left(8\pi\right)} = \frac{\left(\mu_++\mu_-\right)}{\mu_0\pi\mathcal{C}}\left(-2  +\frac{1}{\mathcal{C}}\log\left[\frac{1+\mathcal{C}}{1-\mathcal{C}}\right]\right)~,
\ee
which agrees with the previous result reported in reference [\onlinecite{Mukherjee_2018}].
Note that $\mathcal{Q}_0$ drops out of the dc limit which depends only on $\mu_0$. Eq.~\eqref{xy_dc} shows that the tilt makes an important contribution to ${\rm Re}\left[\sigma_{xy}\left(0\right)\right]$. In the limit of small tilt the leading order in $\mathcal{C}$ is linear and consequently this contribution is only non-zero when there is a tilt.
%
%
%
Note that the expression for real part of the Hall conductivity does not depend on the cut-off in contrast to that of the imaginary part of longitudinal conductivity. This is because of the fact that the imaginary part of Hall conductivity is strictly limited to some particular frequency regions, which are fixed by the chemical potential and tilt parameter [see Fig.~\ref{fig2}(c),(d)]. Moreover, the cut-off dependent terms in the longitudinal part will cancel if we also take into account the diamagnetic response [Ref.~\onlinecite{Steiner}].
From Eq.~\eqref{Hall}, we can see that, $\tan\Theta_H\left(\Omega\right)$
\be\label{Hall_exp}
 = \frac{{\rm Re}\left[\sigma_{xx}\left(\Omega\right)\right]{\rm Re}\left[\sigma_{xy}\left(\Omega\right)\right] + {\rm Im}\left[\sigma_{xx}\left(\Omega\right)\right]{\rm Im}\left[\sigma_{xy}\left(\Omega\right)\right]}{{\left({\rm Re}\left[\sigma_{xx}\left(\Omega\right)\right]\right)}^2 + \left({{\rm Im}\left[\sigma_{xx}\left(\Omega\right)\right]}\right)^2}.
\ee
Note from the structure of the numerator in Eq.~\eqref{Hall_exp} that, because both ${\rm Re}\left[\sigma_{xx}\left(\Omega\right)\right]$ and ${\rm Im}\left[\sigma_{xy}\left(\Omega\right)\right]$ are zero for photon energies less that $2\mu_-/(1+\mathcal{C})$, the Hall angle will be zero as well provided of course $\mu_-$ is non zero. We will return to this point later. Note further that ${\rm Im}\left[\sigma_{xy}\left(\Omega\right)\right]$ has a natural cutoff at energy $2\mu_+/(1-\mathcal{C})$  so that the second contribution to the Hall angle in Eq.~\eqref{Hall_exp} vanishes beyond this point. This is not the case for the first contribution. The real part of $\sigma_{xx}\left(\Omega\right)$ is positive and grows linearly at large $\Omega$ while the real part of Hall conductivity $\left(\text{Fig.}~\ref{fig4}(\text{b})\right)$ becomes negative and small but is not zero. Consequently we expect the Hall angle to get small but remain negative.

In Fig.~\ref{fig4}(a) and Fig.~\ref{fig4}(b) we show results for ${\rm Im}\left[\sigma_{xx}\left(\Omega\right)\right]$ and ${\rm Re}\left[\sigma_{xy}\left(\Omega\right)\right]$ respectively.
Results for  ${\rm Re}\left[\sigma_{xx}\left(\Omega\right)\right]$ and  ${\rm Im}\left[\sigma_{xy}\left(\Omega\right)\right]$ have already been presented in Fig.~\ref{fig2}. In Fig.~\ref{fig4}(a) the green circles for $\mu_p = 0.8\mu_0$ fall close to the black dashed curve for $\mu_p = 0$ (no pumping). Also shown separately are the interband (red curve) and the intraband (blue curve) contribution for $\mu_p = 0.8\mu_0$. In Fig.~\ref{fig4}(b), corresponding results are shown for ${\rm Re}\left[\sigma_{xy}\left(\Omega\right)\right]$. In this case, the differences between green circles and the black dashed curve are more significant, particularly in the region of the maximum and minimum of the black dashed curve. Note that neither of these quantities have a sharp cutoff at small or large $\Omega$ although ${\rm Re}\left[\sigma_{xy}\left(\Omega\right)\right]$ does become small and negative as $\Omega\to\infty$, as we have already commented on. 

\begin{figure}[ht]
\includegraphics[width = 0.9\linewidth]{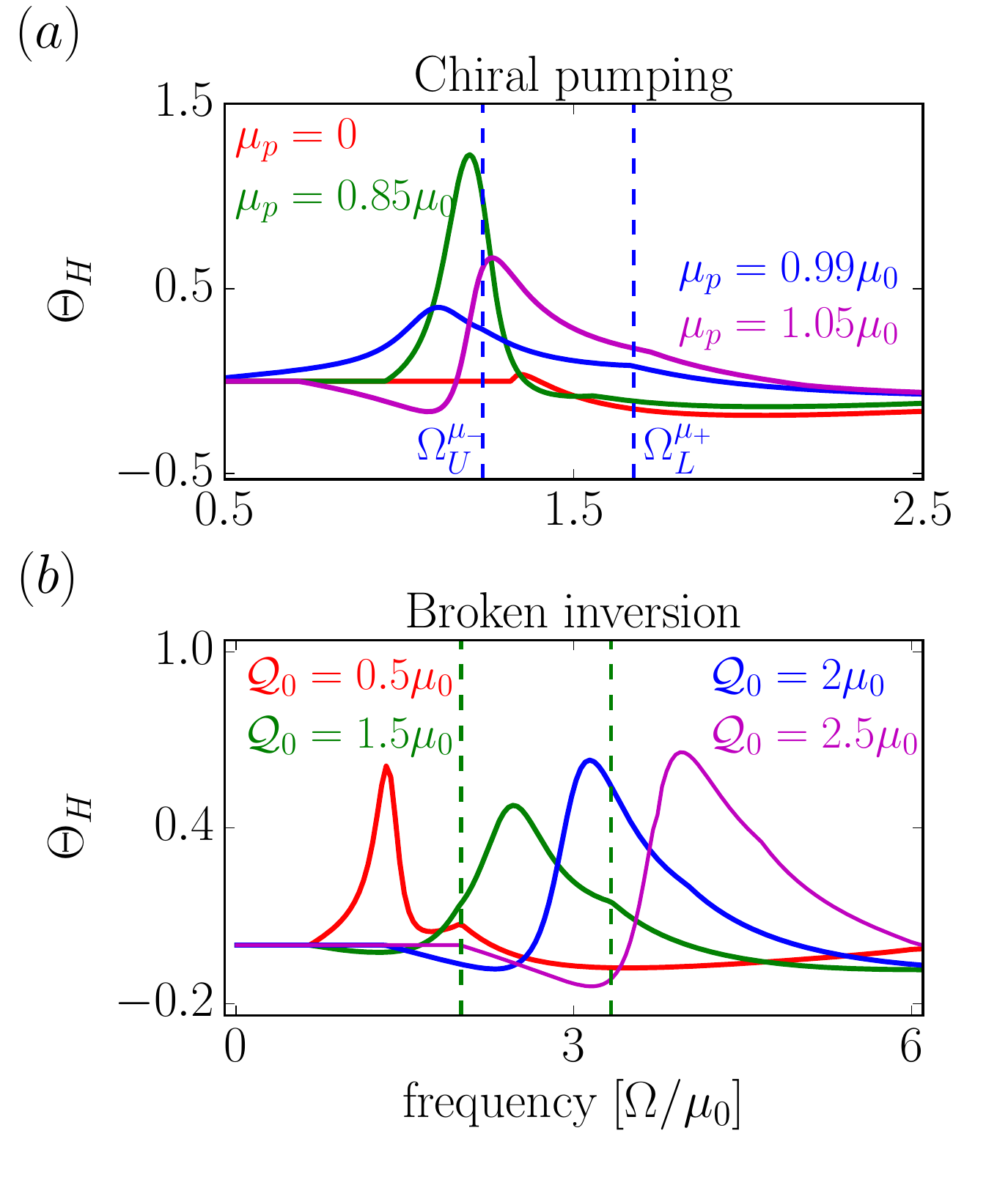} 
\caption{Bulk Hall angle as a function of frequency for (a) the non-vanishing chiral pumping case, where the red curve is the no pumping case ($\mu_p = 0$) and is for comparison, and (b) the broken inversion symmetry case. 
Other parameters are the same as those of Fig. \ref{fig2}.}
\label{fig5}
\end{figure}

Our results for the Hall angle, $\Theta_H~\left(\text{in~radians}\right)$  are presented in Fig.~\ref{fig5}(a) for several values of pumping namely, $\mu_p = 0$~(red), $\mu_p = 0.85\mu_0$~(green),  $\mu_p = 0.99\mu_0$~(blue) and  $\mu_p = 1.05\mu_0$~(indigo). For no pumping, the chemical potential of the two nodes are equal and we observe that the Hall angle is always negative except for a very small positive peak at the onset. 
This peak can be understood by looking at the numerator of Eq.~\eqref{Hall_exp} which has two terms and their sum sets the sign on the Hall angle. Just above the lower energy cut off, the first term is positive because both ${\rm Re}\left[\sigma_{xx}\left(\Omega\right)\right]$ and ${\rm Re}\left[\sigma_{xy}\left(\Omega\right)\right]$ are positive. But ${\rm Re}\left[\sigma_{xy}\left(\Omega\right)\right]$ has a peak in this region (Fig.~\ref{fig4}(b), black dashed curve) and then changes sign. The second term in Eq.~\eqref{Hall_exp} is always negative in the region of interest. Consequently the sum of these two terms can only be positive in a small interval above the lower cutoff, and then become negative. At large $\Omega$ the Hall angle remains negative as it approaches zero (solid red curve in Fig.~\ref{fig5}(a)). The continuous green  curve for $\mu_p = 0.85\mu_0$ behaves in much in the same way as the red but the peak above the onset is now much more pronounced, but in the negative region for the Hall angle, $\Theta_H$,
the two curves track each other well. The solid blue curve is for $\mu_p = 0.99\mu_0$, and shows two small kinks identified in the figure as $\Omega^{\mu_-}_U$ and $\Omega^{\mu_+}_L$. These arise due to the unequal chemical potential in the two nodes. The region between these kinks is one where there is no dichroism (see Fig.~\ref{fig3} panel (c)).
In the Hall angle this manifests as a plateaux type behavior around the end of this frequency interval, as the second kink is approached. The indigo curve is for $\mu_p = 1.05\mu_0$ and shows yet another possible variation for the Hall angle. Here the first peak is negative and then it has a second peak which is positive. This variation is understood as follows. For this case the effective chemical potential of the negative chirality node becomes negative (hole doping). This changes the sign of ${\rm Im}[\sigma_{xy}(\Omega)]$ associated with this node and this changes the sign of the numerator in Eq.~\eqref{Hall_exp} so the first peak is a valley (negative) followed at larger values of $\Omega$ by a positive peak.
%
%

\section{Broken inversion symmetry}\label{Q0}
\begin{figure}[ht]
\includegraphics[width = \linewidth]{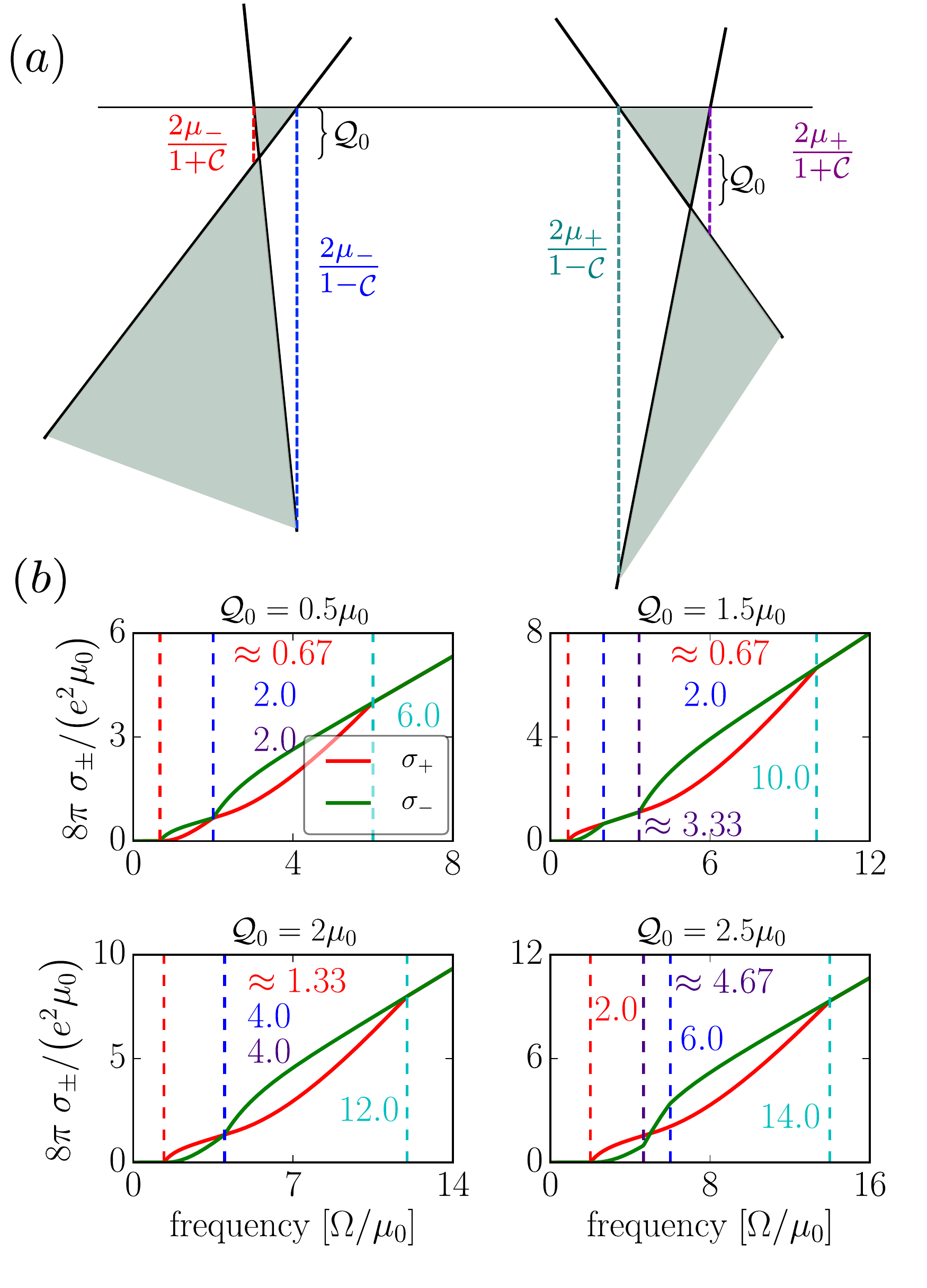} 
\caption{(a) Schematic for various transitions involved when inversion symmetry is broken. (b)Absorptive part of the conductivity for right and left circularly polarized light as a function of frequency for $\mathcal{Q}_0 = 0.5\mu_0$, $\mathcal{Q}_0=1.5\mu_0$, $\mathcal{Q}_0=2\mu_0$ and $\mathcal{Q}_0=2.5\mu_0$. This case is analogous to the chiral pumping case shown in Fig.~\ref{fig3}, except that the red and green curves are less distorted here. Other parameters are the same as those of Fig. \ref{fig2}.}
\label{fig6}
\end{figure}

In this section we consider the case of broken inversion symmetry.
This leads to the energy of the negative chirality node to be pushed up in energy by an amount $\mathcal{Q}_0$ while the positive energy node is pushed down by the same amount. This is illustrated in Fig.~\ref{fig6}(a). 
Our motivation to pursue the present case is to draw analogy from the chiral pumping case. Therefore, we only consider two Weyl nodes assuming the time reversal symmetry is broken. Furthermore, we have assumed the positive and negative chirality nodes to be tilted oppositely as in the chiral anomaly case, although the inversion symmetry in broken in this case. Just as with pumping, there is a different effective chemical potential in the two Weyl cones and the formulas developed so far can still be used. Differences will arise only from the different relationship between
the two chemical potentials involved.
%

In Fig.~\ref{fig6}(b), we show results for the absorptive part of circularity polarized 
light $\left(\sigma_{\pm}\right)$. Four values of $\mathcal{Q}_0$ have been considered, namely,  $0.5\mu_0$, $1.5\mu_0$, $2\mu_0$ and $2.5\mu_0$. In real materials\cite{Lv2015}, the energy separation between Weyl points ranges in $10-20$ meV. By taking $\mu_0 = 10$ meV, we can see that the ratio, $\mathcal{Q}_0/\mu_0$ is adequately represented by our choice of parameters.
Note that $\mathcal{Q}_0 = 0$ and $\mathcal{Q}_0 = \mu_0$ are identical to case of $\mu_p = 0$ and $\mu_p = \mu_0$ respectively.
In the first frame on the left, the red curve for $\sigma_+$ is always below the green curve for $\sigma_-$ because the imaginary part of the Hall conductivity is positive for both nodes. There is no absorption below photon energy of $2\mu_0/3$ (first vertical red dashed line). Above this energy till we reach a photon energy of $2\mu_0$, the negative chirality node has a non zero imaginary part of the dynamic Hall conductivity
and it is only this node that contributes to the dichroism and indeed to the absorption. The positive chirality node contributes only above $\Omega = 2\mu_0$. While in this energy range both nodes contribute to the absorption, only the positive chirality contributes to the dichroism. Beyond $\Omega = 6\mu_0$, there is no dichroism and both nodes contribute equally to the absorption and this dependence on $\Omega$ is that of the no doping, no tilting case. In the second frame for $\mathcal{Q}_0 = 1.5\mu_0$, there is a region at small energy where there is no absorption followed by a second region where the red curve is now above the green. 
This feature is due to the fact that the effective chemical potential in the negative chirality node is negative (hole doping) and this changes the sign of ${\rm Im}[\sigma_{xy}(\Omega)]$. At yet higher photon energies there is a region
 of no dichroism and the absorption is only from the negative chirality node. Above this energy the positive chirality node kicks in and red and green curves switch. The dichroism is now due only to this node. Above the fourth vertical line, which marks the end of a finite Hall conductivity there is no dichroism. The other two frames serve to show the richness of behaviors that can arise.

For $\mathcal{Q}_0 = 2\mu_0$, in the region between first and second horizontal lines only the negative chirality node is involved and its effective chemical potential is negative so the green and red curves switch. In the third interval the positive chirality node has kicked in while the negative
chirality node has no Hall contribution. Consequently in this region the dichroism is entirely due to the positive chirality node.
The final value chosen for $\mathcal{Q}_0 = 2.5\mu_0$ shows yet another different behavior.
 In the interval between the first two vertical lines, we have the same behavior as in the case $\mathcal{Q}_0 = 2\mu_0$ but beyond that there is an important difference because the contribution of the negative chirality node now extends to the third vertical line and the second region involves a finite imaginary Hall contribution from both nodes with opposite signs (negative for negative chirality node and positive for positive chirality node).
 This is why the crossing point between $\sigma_{+}$ and $\sigma_{-}$ does not fall on one of the vertical lines but is determined by the crossing of negative chirality node and positive chirality node Hall conductivity which have the opposite sign. In the fourth interval only the positive chirality node makes a contribution to the dichroism.
\begin{figure}[ht]
\includegraphics[width = \linewidth]{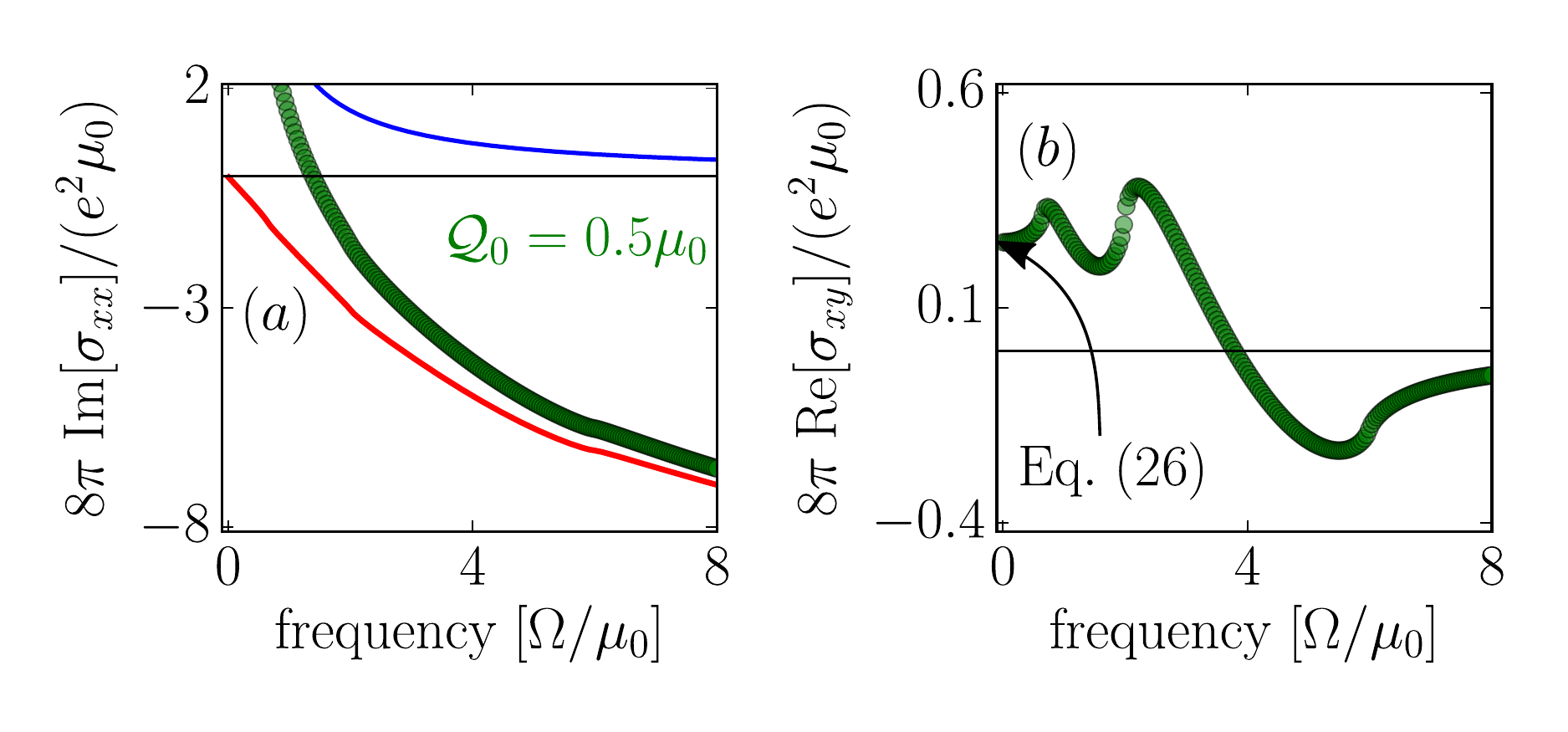} 
\caption{(a) The imaginary part of longitudinal and (b) real part of the Hall conductivity is shown as a function of frequency for broken inversion symmetry case. The bubbled line in (a) refers to the total sum of the contribution from the inter and the intra-band conductivity for $\mathcal{Q}_0 = 0.5\mu_0$. In panel (b), ${\rm Re}\left[\sigma_{xy}\left(\Omega\right)\right]$ is shown for the same value of $\mathcal{Q}_0$. Note that its dc value is consistent with Eq.~\eqref{xy_dc}. Other parameters are the same as those of Fig. \ref{fig2}.}
\label{fig7}
\end{figure}

In Fig.~\ref{fig7} we show results for the imaginary part of longitudinal conductivity in frame (a) and real part of Hall conductivity in frame (b), as in Fig.~\ref{fig4} but now we have $\mathcal{Q}_0 = 0.5\mu_0$ (green circles) compared with $\mathcal{Q}_0 = 0$ case (see Fig.~\ref{fig4}(b)). As Eq.~\eqref{xy_dc} tells us, $\mathcal{Q}_0$ drops out of the dc limit for ${\rm Re}\left[\sigma_{xy}\left(\Omega\right)\right]$ but introduces new structure as a function of $\Omega$ when compared with frame (b) of Fig.~\ref{fig4} and this leads to important differences in the behavior of the Hall angle.

The results for Hall angle are shown in Fig.~\ref{fig5}(b) for four values of $\mathcal{Q}_0$, namely, $\mathcal{Q}_0 = 0.5\mu_0$ (red), $\mathcal{Q}_0 = 1.5\mu_0$ (green), $\mathcal{Q}_0 = 2\mu_0$ (blue) and $\mathcal{Q}_0 = 2.5\mu_0$ (indigo). Only the first (red) curve corresponds to electron doping for both nodes. In the other cases the doping in the negative chirality node involve holes and as expected all these curves show, first a valley followed by a peak at higher energies as found for the chiral pumping case with $\mu_p = 1.05\mu_0$. The characteristics of the curves are similar to those already described in detail for the pump case and will not be emphasized further here. We point out one feature namely that, for the green curve for which there is a region of no dichroism between second and third vertical line in Fig.~\ref{fig6}(b) (right top frame), we again see a plateau type structure ending in a kink as we emphasized in the blue curve in Fig.~\ref{fig5}(a).
\section{Summary}\label{summ}
In this work we have considered the optical properties of a doped and tilted Weyl semimetal. With tilting and a finite chemical potential, such systems
acquire a finite transverse dynamic Hall conductivity in a photon energy interval between  $2\mu_0/\left(1+\mathcal{C}\right)$ and $2\mu_0/\left(1-\mathcal{C}\right)$. This leads to dichroism for the absorption of circular polarized light. When a parallel electric and magnetic field is applied to the sample, there is a charge transfer from the negative to the positive chirality node due to the chiral anomaly. This leads to a new steady state in which the effective chemical potential of the negative chirality node is smaller than $\mu_0$ and that in the positive chirality node is larger. Another way of producing a
difference in the effective chemical potential between the nodes is to break inversion symmetry and consider the noncentro symmetric case. In both these instances the dichroism is modified as is the dynamic Hall angle. Here we have considered both cases and have provided analytic results for
both the real and imaginary part of the dynamic longitudinal and  transverse Hall conductivity from which  both the dichroism and Hall angle follow. We have found a rich pattern of possible behaviors depending on the size of the differences in effective chemical potential of the two nodes. 

When considering absorption, there can be regions of photon energies in which only one of the nodes contributes to the longitudinal conductivity, and others where both contribute. This is also the case for the imaginary part of the Hall conductivity. A case considered
in detail shows a first region at low energy for which there is no absorption of both right and left handed circularly polarized light, followed by a region to which only the negative chirality node contributes, but there is nevertheless dichroism. This is followed by another region which still is 
a result of only the negative chirality node but now displays no dichroism
and the absorptive part of the conductivity is linear in photon energy with slope $1/3$ in our reduced units. At still higher energies both nodes contribute to the absorption but only the positive chirality node is involved in dichroism. Finally there is a region at higher energies where the well known, no doping no tilting behavior of a Weyl semimetal is recovered where the slope of the linear behavior is equal to $2/3$. Other interesting patterns also emerge. 

The dynamic Hall angle is strongly modified by pumping due to the chiral anomaly or due to broken inversion symmetry. As an example for no pumping and inversion symmetry, the Hall angle is negative for photon energies between $2\mu_0/\left(1+\mathcal{C}\right)$ and $2\mu_0/\left(1-\mathcal{C}\right)$ and zero everywhere else except for a small peak at its onset. With pumping this all changes, although for small $\mu_p$ values, the changes from its zero pumping value are small except that the initial peak can be much larger. When $\mu_p = 0.99\mu_0$, there is a positive peak above the onset energy and then a slower decay towards zero Hall angle as $\Omega\to\infty$. There is also a region where the Hall angle plateaux and shows relatively little change with changing $\Omega$. This region corresponds to a region of frequency where there is no dichroism. For $\mu_p = 1.05\mu_0$, the behavior is similar but with an initial negative dip before a large positive peak is seen. The reason for a dip rather than a peak is traced to the contribution from the negative chirality node which is hole doped. A second rich pattern of possible behaviors for the Hall angle is predicted for the case of no pumping but broken inversion symmetry.

A limitation to this work is the consideration that the measurements we propose would be carried out in clean samples and at low temperatures such that the effect of impurities and thermal broadening can be ignored. This is of course a difficult aim to achieve in real experiments. However we believe that some of the features can still be observed by tuning few parameters, e.g. the value of applied electric or magnetic field in case of chiral pumping. We also expect that the dichroism displayed in the noncentrosymmetric case may not vanish completely for cases when $\mathcal{Q}_0 > \mu_0$ as long as thermal energy is less that $\mu_0$.
\section{Acknowledgement}
Work supported in part by the National Science and Engineering Research Council of Canada (NSERC) and by the Canadian Institute for Advanced Research (CIFAR) (Canada). We thank T. Timusk, E. J. Nicol and S. Mardanya for enlightening discussions.
\bibliography{Ref2}

\end{document}